\documentclass[twocolumn,superscriptaddress,longbibliography,
	aps,pra,preprintnumbers]{revtex4-2}

\usepackage[normalem]{ulem}
\usepackage{graphicx}
\usepackage{bm}
\usepackage{color}
\usepackage{epstopdf}
\usepackage{amsmath}
\usepackage{amssymb}
\usepackage{epstopdf}
\usepackage{lipsum}

\usepackage{gensymb}

\usepackage{multirow}

\usepackage{scrextend}

\usepackage[urlcolor=blue,colorlinks=true,citecolor=blue,linkcolor=blue,pdfstartview={FitH},bookmarks=false]{hyperref}

\usepackage[dvipsnames]{xcolor}
\definecolor{mygreen}{rgb}{0.0, 0.6, 0.0}
\definecolor{pjorange}{rgb}{0.8, 0.3, 0.0}
\definecolor{jlblue}{rgb}{0.2, 0.5, 0.7}

\graphicspath{{fig/}{./fig/}{.}}

\begin{document}

\title{
Ruthenium dioxide RuO$_{2}$: effect of the altermagnetism on the physical properties
}

\author{Andrzej~Ptok}
\email[e-mail: ]{aptok@mmj.pl}
\affiliation{\mbox{Institute of Nuclear Physics, Polish Academy of Sciences, W. E. Radzikowskiego 152, PL-31342 Krak\'{o}w, Poland}}

\date{\today}

\begin{abstract}
Ruthenium oxide with the rutile structure is one of example of altermagnets. 
These systems are characterized by compensated magnetic moments (typical for antiferromagnets) and strong time reversal symmetry breaking (typical for ferromagnets).
However, in such cases, the electronic band structure exhibit strong spin splitting along some directions in the momentum space.
Occurrence of the compensated magnetic textures allows for realization of surfaces with specific magnetization, which dependent on the surface orientation and/or its termination. 
Here, we study interplay between the electronic surface states and the surface magnetization.
We show that the spin-resolved spectra strongly depends on a direction in reciprocal space.
Such properties can be used for the experimental confirmation of the altermagnetism in RuO$_{2}$ within the spectroscopic techniques.
Additionally, we show that the most modified orbitals in the system are $d_{z^{2}}$ and $d_{xy}$ orbitals of Ru.
Similarly, the Ru $e_{g}$ states are most sensitive on epitaxial strain, what can suggest some link between altermagnetism and strain.
\end{abstract}

\maketitle

\begin{figure}[!b]
\centering
\includegraphics[width=\columnwidth]{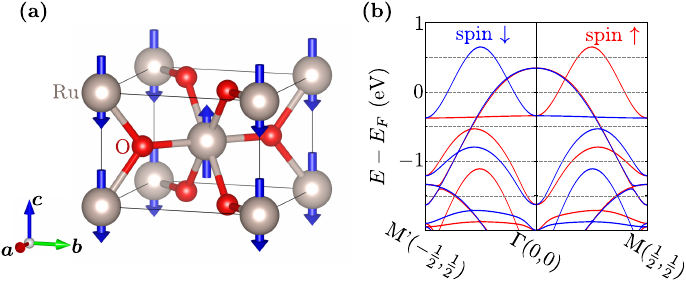}
\caption{
(a) The crystal structure of RuO$_{2}$ with the rutile structure.
(b) The electronic band structure, showing a strongly broken time reversal symmetry and altermagnetic features of RuO$_{2}$.
\label{fig.schemat}
}
\end{figure}

\section{Introduction}
\label{sec.intro}

In the simplest case of crystals with collinear magnetic order, the ferromagnetic and antiferromagnetic systems were considered.
The ferromagnetism corresponds to the strong magnetization and to the spin-polarization in electronic bands that breaks the time reversal symmetry.
In contrast to this, the antiferromagnetism generates compensated collinear order with no total magnetization.
More recently, altermagnets -- a third elementary class of crystal with a collinear magnetic order has been discovered~\cite{smejkal.sinova.22,smejkal.sinova.22b}.
In such a case, the system exhibits a symmetry-protected compensated antiparallel magnetic order on a crystal lattice that generates unconventional alternating spin-polarization and the time-reversal-symmetry breaking in the band structure without magnetization~\cite{smejkal.gonzalezhernandez.20}.

One of compounds exhibiting the altermagnetism is RuO$_{2}$ with the rutile structure (Fig.~\ref{fig.schemat}).
In this material, the collinear antiferromagnetic order is observed in relatively high temperatures ($T_{N} > 300$~K)~\cite{berlijn.snijders.17,zhu.strempfer.19}.
Combination of the roto-reversal symmetry~\cite{gopalan.litvin.11} and compensated antiparallel magnetic moments leads to the anomalous Hall effect of comparable strength 
as for other ferromagnets~\cite{feng.zhou.22,smejkal.macdonald.22}.
As a consequence, the chiral magnons can be realized~\cite{smejkal.marmodoro.22}, what has a potential to application in the novel spinotronic devices~\cite{gonzalez.smejkal.21,bose.schreiber.22,shao.zhang.21,smejkal.hellenes.22}.
Additionally, this compound exhibits several interesting features like
non-trivial topological properties~\cite{sun.zhang.17,jovic.koch.18,zhan.li.23}, and superconductivity induced by epitaxial strain~\cite{uchida.nonoto.20,ruf.paik.21}.
Moreover, as a highly conductivity binary oxide~\cite{ryden.lawson.68,ryden.lawson.70}, it could be used in many engineering applications~\cite{music.kremer.15}.

Realization of the altermagnetism in RuO$_{2}$ is associated with strong time-reversal symmetry breaking, which is exhibited by strong mismatch between opposite spin bands along some direction [Fig.~\ref{fig.schemat}(b)].
Nevertheless, in real space, the system is characterized by compensated magnetic moments [Fig.~\ref{fig.schemat}(a)], what allows for a realization of the surfaces terminated by Ru atoms belonging to the same magnetic sublattice and with the same magnetization. 
Additionally, we can expect that the surface orientation can affect the physical properties of the system.
Indeed, for example, the spin orientation on the surface affects on the thermal conductivity of the material~\cite{zhou.feng.23}. 
The role of the RuO$_{2}$ surface magnetism is also well know in the context of catalyses~\cite{torun.fang.13} or oxygen evolution reaction~\cite{liang.bieberlehutter.22}.
More recently, also the strong time-reversal symmetry breaking was observed in magnetic circular dichroism~\cite{fedchenko.minar.23}, as well as on spin-dependent emitted light experiments~\cite{liu.bai.23}.
However, the problem of the interplay between altermagnetism and electronic properties is still open, especially in the context of topological features of RuO$_{2}$-like compounds~\cite{sun.zhang.17,jovic.koch.18,zhan.li.23,nelson.ruf.19}, or generally antiferromagnetic compounds~\cite{bonbien.zhuo.22}.

In this paper, we discuss consequences of occurring altermagnetism on the surface states dependent on the surface orientation and termination. 
Recent study indicate that the spin-resolved spectroscopy can be used for confirmation of the altermagnetism in MnTe~\cite{lee.lee.23}.
Indeed, spin splitting can affect on the spin polarization of the surface states.
However, mentioned MnTe material~\cite{lee.lee.23} is characterized by relatively small band splitting. 
Contrary to this, in the RuO$_{2}$ band splitting is around $1$~eV~\cite{smejkal.sinova.22b} and is much larger than in other altermagnets~\cite{guo.liu.23}.
This makes RuO$_{2}$ as an excellent candidate for experimental study of the altermagnetism within the spectroscopy techniques~\cite{king.picozzi.21}.

The paper is organized as follows.
The computational details are present in Sec~\ref{sec.comp}.
Next, in Sec.~\ref{sec.res}, we present and discuss our numerical results.
Finally, a summary is included in Sec.~\ref{sec.sum}.

\begin{figure}[!b]
\centering
\includegraphics[width=\columnwidth]{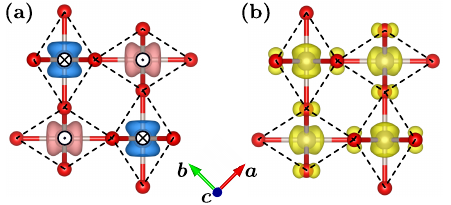}
\caption{
Realization of the roto-reversal symmetry in RuO$_{2}$ and its consequences on the spin (a) and charge (b) density from {\it ab initio} calculations.
\label{fig.roto_reversal}
}
\end{figure}

\section{Computational details}
\label{sec.comp}

The first-principles density functional theory (DFT) calculations were performed using the projector augmented-wave (PAW) potentials~\cite{blochl.94} implemented in the Vienna Ab initio Simulation Package ({\sc Vasp}) code~\cite{kresse.hafner.94,kresse.furthmuller.96,kresse.joubert.99}.
For the exchange-correlation energy the generalized gradient approximation (GGA) in the Perdew, Burke, and Ernzerhof (PBE) parametrization was used~\cite{pardew.burke.96}.
Similarly to the previous study, we introduce the correlation effects on Ru $d$ orbitals within DFT+U approach, introduced by Dudarev {\it et al.}~\cite{dudarev.botton.98}.
The energy cutoff for the plane-wave expansion was set to $600$~eV.
Optimization of the structural parameters (in the presence of the spin--orbit coupling) were performed using $10 \times 10 \times 15$ ${\bm k}$--point grid using the Monkhorst--Pack scheme~\cite{monkhorst.pack.76}.
As a convergence condition of the optimization loop, we took the energy change below $10^{-6}$~eV and $10^{-8}$~eV for ionic and electronic degrees of freedom.

The study of the electronic surface properties we construct the tight-binding model in the maximally localized Wannier orbitals~\cite{marzari.vanderbilt.97,souza.marzari.01,marzari.mostofi.12}.
In this case, the DFT calculation were performed within {\sc Quantum ESPRESSO}~\cite{giannozzi.baseggio.20}.
Calculations were done on $8 \times 8 \times 8$ ${\bm k}$-grid, within DFT+U scheme (like previously) and PBE PAW pseudopotentials developed within {\sc PSlibrary}~\cite{dalcorso.14}.
Here, we take $100$~Ry and $450$~Ry for wavefunctions and charge density cutoffs, respectively.
The exact band structure was used to construct the tight-binding model by {\sc Wannier90}~\cite{pizzi.valerio.20}, based on Ru $d$ and O $p$ orbitals, what correspond to a $44$-band model.
Finally, the electronic surface states were calculated using the surface Green's function technique for a semi-infinite system~\cite{sancho.sancho.85}, implemented in {\sc WannierTools}~\cite{wu.zhang.18}.


\section{Results and discussion}
\label{sec.res}

\subsection{Crystal structure}

RuO$_{2}$ crystallizes with the rutile structure (P4$_2$/mnm, space group No.~136) presented in Fig.~\ref{fig.schemat}(a).
In this case, the Ru and O atoms are located in Wyckoff positions $2b$ (0,0,0) and $4g$ ($x_{O}$,$y_{O}$,0), respectively.
Here, $x_{O}$ and $y_{O}$ are two free parameters describing position of O atoms in the crystal structure.

The obtained crystal parameters for different $U$ are collected in Table~\ref{tab.latt} in the Supplementary Materials (SM)~\footnote{See Supplemental Material at [URL will be inserted by publisher] for additional numerical results, containing the lattice parameters for different strength interaction $U$, the electronic band structure for different magnetic moment directions, and discussion of the impact of epitaxial strain on the electronic density of states.}.
In practice, all lattice parameters are similar, and only Ru magnetic moments strongly depends by $U$.
Similarly like in the previous study~\cite{ahn.hariki.19}, we assume $U = 2$~eV, for which Ru posses the magnetic moment $1.152$~$\mu_{B}$.
In this case, $a = 4.533$~\AA, $c = 3.11$~\AA, while the free parameters are obtained as $x_{O} = 0.8037$ and $y_{O} = 0.1963$. 
The lattice constant are closed to the experimentally observed ones, i.e., $a = 4.49$~\AA, and $c = 3.11$~\AA~\cite{zhu.strempfer.19}.

\begin{figure}[!b]
\centering
\includegraphics[width=\columnwidth]{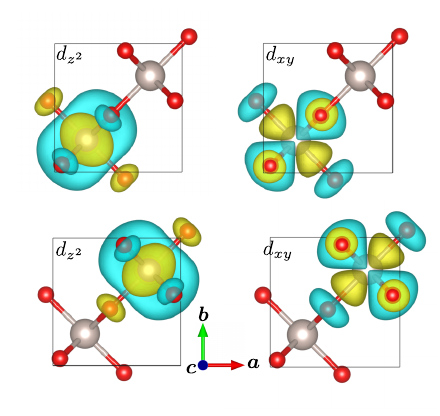}
\caption{
The strongest modified Wannier orbitals in Ruo$_{2}$ altermagnet: $d_{z^{2}}$-like and $d_{xy}$-like molecular orbitals (left and right panels, respectively).
Color of isosurface corresponds to the wave-function phase.
\label{fig.mol_orb}
}
\end{figure}

\subsection{Magnetic moments and spin--orbit coupling}

RuO$_{2}$ contains two sublattices with different Ru magnetic moment orientation [see Fig.~\ref{fig.schemat}(a)].
A relation between these two sublattices is described by the screw four-folded rotational symmetry $\{ 4^{+}_{001} | \frac{1}{2} \frac{1}{2} \frac{1}{2} \}$ around $z$ axis, which is one of space group P4$_{2}$/mnm generators.
This generator acting on the one Ru sublattice transforms it to the second sublattice with opposite Ru magnetic moments [see Fig.~\ref{fig.roto_reversal}(a)].
Moreover, this symmetry is also ``inherited'' by the spin/charge density.
This is related to the relation between Ru $d$ orbitals in both sublattices given by the mentioned space group generator.
Indeed, the direct calculations of the spin and charge densities within {\it ab initio} exhibit the same features (see Fig.~\ref{fig.roto_reversal}).
However, analyses of the Wannier orbitals realized in the system reveals that the strongest modified orbitals have form of $d_{z^{2}}$-like and $d_{xy}$-like orbitals (Fig.~\ref{fig.mol_orb}).

\begin{figure}[!b]
\centering
\includegraphics[width=\columnwidth]{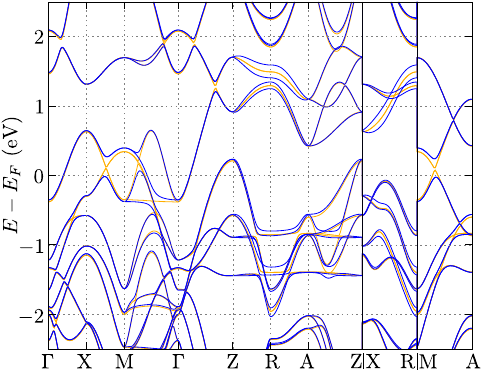}
\caption{
Electronic band structure of RuO$_{2}$ in the absence and in the presence of the spin--orbit coupling (orange and blue lines, respectively).
\label{fig.el_band}
}
\end{figure}

There is also well-know sensitivity of RuO$_{2}$ on epitaxial strain~\cite{uchida.nonoto.20,ruf.paik.21}, which can induce the superconductivity in this compound.
To check this behaviour, we introduce the intentional modification of $a = b$ lattice constants, and found corresponding $c$ lattice parameter, which minimize the system energy (see Fig.~\ref{fig.el_strain_dos} in SM~\cite{Note1}).
This should lead to mimic the effect of epitaxial strain, due to the mismatch between such modified RuO$_{2}$ and based material in $ab$ plane. 
For this fixed parameters, we obtain electronic density of states.
The electronic density of states analyses reveals strongly sensitive Ru $e_{g}$ orbitals on the epitaxial strain, what is in an agreement with the previous study~\cite{gregory.strempfer.22}.
This is interesting in the context of the molecular orbitals realized in RuO$_{2}$.
The strongest modified Ru $d$ orbitals take form of $d_{z^{2}}$-like and $d_{xy}$-like orbitals (Fig.~\ref{fig.mol_orb}), even if the band crossing at the Fermi level have Ru $d_{xz}$ and $d_{yz}$ character~\cite{guo.liu.23}.
This can suggest a link between epitaxial strain and altermagnetism.

\begin{figure}[!t]
\centering
\includegraphics[width=0.75\columnwidth]{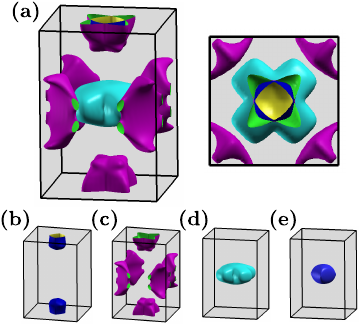}
\caption{
The Fermi surface of RuO$_{2}$ in the presence of spin--orbit coupling.
Top panels (a) present the full Fermi surface, while bottom panels (b)--(e) present the separate Fermi pockets.
\label{fig.el_fermi_soc}
}
\end{figure}

\begin{figure}[!b]
\centering
\includegraphics[width=\columnwidth]{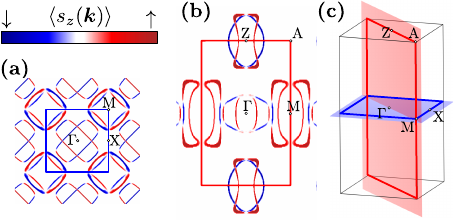}
\caption{
Spin polarization of the bulk Fermi surface (a,b) for two different cross-section of the Brillouin zone (c).
\label{fig.el_spin_fermi}
}
\end{figure}

\begin{figure*}
\centering
\includegraphics[width=\textwidth]{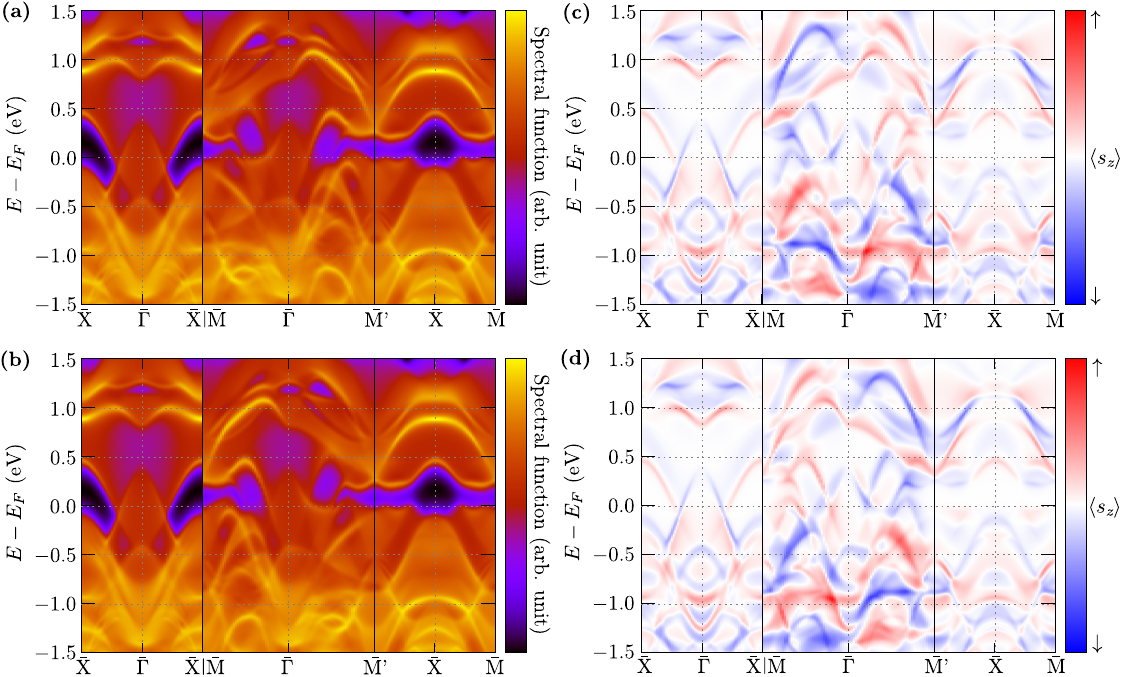}
\caption{
Spectral function (left column) and spin polarization (right column) of the states for (001) surface, terminated by Ru atoms with $\uparrow$ and $\downarrow$ magnetic moments (upper and lower row, respectively).
Here, $\bar{M} = (\frac{1}{2} \frac{1}{2})$ and $\bar{M'} = (-\frac{1}{2} \frac{1}{2})$.
\label{fig.sp001}
}
\end{figure*}

The magnetic moments on Ru atoms exhibit an ordinary anitferromagnetic order [Fig.~\ref{fig.schemat}(a)].
However, direction of the magnetic moments on Ru is still under debate~\cite{lovesey.khalyavin.23}.
In our calculations, the system energy is minimized when the Ru magnetic moments are parallel to $c$ axis.
Additionally, the value of Ru magnetic moment weakly depends on a assumed direction and it changes in range of $1$~$\mu_{B}$ (for $U = 2$~eV).
Moreover, the magnetic moment direction has negligible impact on the electronic band structure [cf. Fig.~\ref{fig.el_band} and Fig.~\ref{fig.el_band_tilted} in SM~\cite{Note1}].
As a consequence of the changed  magnetic moment direction, only additional tiny (invisible on the figures) band splitting appears.

The electronic band structure in the absence of the spin--orbit coupling is presented on Fig.~\ref{fig.schemat}(b).
The spin-splitting of the band observed along two perpendicular $\Gamma$--M paths exhibits alternating structure (cf. red and blue color bands).
Due to the crystal symmetry, each Ru atoms is located inside octahedrons formed by O atoms.
As a consequence, the band splitting originate only from the different magnetic moments in Ru sublattices.
As we can see, the band splitting is around $1$~eV. 
Introduction of the spin--orbit coupling leads to opening a gap at the place of the band crossing (Fig.~\ref{fig.el_band}).
The spin--orbit strength is the biggest one on Ru atoms (around $3$~meV and $11$~meV for $p$ and $d$ orbitals, respectively), whereas for O atoms is negligible small (around $0.05$~meV for $p$ orbitals).
The magnetic anisotropy energy (MAE) between states with magnetic moment along $c$ and $a$ is around $0.6$~meV/f.u.

The spin--orbit coupling affects also on the Fermi surfce states.
In the absence of the spin--orbit coupling, the Fermi surface presented on Fig.~\ref{fig.el_fermi} in SM~\cite{Note1} exhibits a toroidal-like deformation~\cite{hayami.yanagi.20} due to the symmetric spin splitting~\cite{hayami.yatsushiro.18}.
The gap induced by the spin--orbit coupling plays a crucial role on the Fermi surface shape [Fig.~\ref{fig.el_fermi_soc}(a)].
In this case, the Fermi surface is constructed by four Fermi pockets [Fig.~\ref{fig.el_fermi_soc}(b)--(e)] with three dimensional character.

As we know, the strength of the spin--orbit coupling strongly depends on the atomic mass~\cite{shanavas.popovic.14}.
RuO$_{2}$ is composed by relatively light atoms, however, the modification of the electronic band structure and the Fermi surface by the relativistic effect in the frame of spin--orbit coupling cannot be neglected.


\subsection{Spin polarization of bulk and surface states}

Independently from the spin--orbit coupling, the band possess strong spin polarization.
Indeed, this is well visible within the spin polarization analyses of the Fermi surface.
In Fig.~\ref{fig.el_spin_fermi}, we present average projection of spin on $z$ axis ($\langle s_{z} \rangle$).
In the case of $\Gamma$--X--M plane ($k_{z} = 0$), the inner Fermi pockets (around $\Gamma$ point) are characterized by $d-wave$ spin--momentum locking~\cite{smejkal.sinova.22}.
In the case of the outer Fermi pockets (around M point,) the spin--momentum locking can be more complicated.

Bands avoiding crossing, due to the gap induced by the spin--orbit coupling, lead to existence of the separable Fermi pockets (see Fig.~\ref{fig.el_fermi_soc}).
The gap induced by the spin--orbit coupling plays important role for occurrence of the topological properties.
This can be shown by the Berry curvature $\Omega ( {\bm k} )$ calculations~\cite{smejkal.gonzalezhernandez.20,zhou.feng.23}.
The non-vanishing integrated Berry curvature gives rise to the intrinsic Hall conductivity.
Nevertheless, even for  specific Fermi pocket, the spin polarization of  the Fermi surface changes significantly (i.e. spin-up $\rightleftarrows$ spin-down, under rotation around $z$ direction).
This means that the nodal lines of the spin polarization (i.e., $\langle s_{z} \rangle = 0$) can be found.

\begin{figure}[!t]
\centering
\includegraphics[width=\columnwidth]{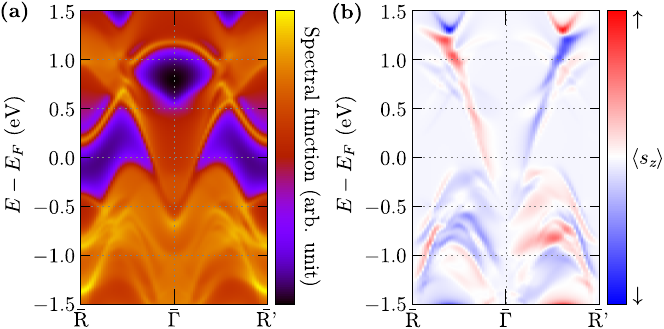}
\caption{
The same as Fig.~\ref{fig.sp001}, but for (100) surface terminated by Ru with $\downarrow$ magnetic moment.
Here $\bar{R} = (\frac{1}{2} \frac{1}{2})$, and $\bar{R'} = (-\frac{1}{2} \frac{1}{2})$.
\label{fig.sp100}
}
\end{figure}

That strong spin polarization of the bulk states can be reflected also within the spin polarization of surface states.
In similar way as above, using semi-infinite method, we can calculate the spin polarization of the surface states as:
\begin{eqnarray}
\langle s_{z} \rangle \left( {\bm k}_{\parallel} , \omega \right) = - \frac{1}{\pi} \lim_{\eta \rightarrow 0^{+}} \frac{ \sigma_{z} G_{s} \left( {\bm k}_{\parallel} , \omega + i \eta \right) }{ A \left( {\bm k}_{\parallel} , \omega \right) } ,
\end{eqnarray}
where $G$ denotes surface Green function, while $A$ is surface spectral function for energy $\omega$ and ${\bm k}$ is parallel to the surface.
In our analyses, we present results for (001) and (100) surfaces (Fig.~\ref{fig.sp001} and~\ref{fig.sp100}, respectively).
For such surfaces, the termination is realized by the Ru atoms from the same magnetic sublattice.
Contrary to this, the (111) surface is terminated by Ru atoms from both sublattices and presented analyses have no application to this surface.

First of all, we should notice, that the (001) and (100) surfaces posses the two-folded rotational symmetry (around normal vector $\hat{\bm n}$ to the surface), what is related with environment of Ru atoms.
This is reflected in the form of ordinary surface states along $\bar{\Gamma}$--$\bar{\text{M}}$ [for (001)] or $\bar{\Gamma}$--$\bar{\text{R}}$ [for (100)], presented in the central part of Fig.~\ref{fig.sp001}(a,b) and Fig.~\ref{fig.sp100}(a), respectively.
In such cases, the surface spectrum function does not possess the symmetry with respect to the two--folded rotational symmetry around $\hat{\bm n}$.
Contrary to this, another directions, like $\Gamma$--$\bar{\text{X}}$ and $\bar{\text{X}}$--$\bar{\text{M}}$ for (001), do not exhibit these features.
Such absence of the surface states symmetry is also reflected in the spin polarization [see Fig.~\ref{fig.sp001}(c,d) and Fig.~\ref{fig.sp100}(b)].
The projection of the strongly spin polarized bulk state on the two dimensional surface Brillouin zone leads to arise the spin polarized surface states.
As we can see, depending on the direction of measurements, the spin polarization possesses the opposite spin polarization.
Similarly, the spin polarization of the Fermi surface states within the constant energy contour measurement should be related to the bulk polarization presented in Fig.~\ref{fig.el_spin_fermi}.
These features should be possible to measure in a relatively simple way  within the spin-polarized spectroscopic techniques~\cite{king.picozzi.21}.

\section{Summary}
\label{sec.sum}

Confirmation of the time-reversal symmetry breaking~\cite{fedchenko.minar.23} and altermagnetism~\cite{lee.lee.23} is a new challenge nowadays. 
In this paper, we discussed the effect of the altermagnetism on the physical properties of ruthenium dioxide RuO$_{2}$, in the context of the spin-resolved spectroscopy.

First of all, we discussed the basic properties of the system, claiming that the most modified Wannier orbitals are Ru $d_{z^{2}}$ and $d_{xy}$ orbitals what is in contrary to the previous study.
Additionally, we show that the Ru $e_{g}$ orbitals are very sensitive to the epitaxial strain, what can also suggest a link between strain and the altermagnetism.
We showed a negligible role of the magnetic moments orientation on the electronic band structure.

Finally, we focused also on the spin-resolved spectroscopic evidence of the spin-polarized (bulk and surface) states.
RuO$_{2}$ is characterized by relatively large band spin splitting, what makes this compound excellent candidate to the altermagnetism study within spectroscopic techniques.
The spin-splitting has a crucial role on the surface orientation with termination by the Ru atoms from one magnetic sublattice.
We showed that the surface states in such case [e.g., (001) or (100) surfaces for RuO$_{2}$] are characterized by the absence of the symmetry in spin-polarized spectra along some directions.
This can be used as a experimental evidence of the altermagnetism in the compounds with large bands spin-splitting.

\begin{acknowledgments}
The author thanks Konrad J. Kapcia for helpful discussions and critical reading of the manuscript. 
Some figures in this work were rendered using {\sc Vesta}~\cite{momma.izumi.11} and {\sc XCrySDen}~\cite{kokalj.99} software.
A.P. is grateful to Laboratoire de Physique des Solides in Orsay (CNRS, University Paris Saclay) for hospitality during a part of the
work on this project.
We kindly acknowledge support by National Science Centre (NCN, Poland) 
under Project No.~2021/43/B/ST3/02166.
\end{acknowledgments}


\bibliography{biblio.bib}


\clearpage
\newpage

\onecolumngrid

\begin{center}
  \textbf{\Large Supplemental Material} \\[.3cm]
  \textbf{\large Ruthenium dioxide RuO$_{2}$: effect of the altermagnetism on the physical properties} \\[.3cm]
  Andrzej Ptok \\[.2cm]
(Dated: \today)
\\[1cm]
\end{center}

\setcounter{equation}{0}
\renewcommand{\theequation}{SE\arabic{equation}}
\setcounter{figure}{0}
\renewcommand{\thefigure}{SF\arabic{figure}}
\setcounter{section}{0}
\renewcommand{\thesection}{SS\arabic{section}}
\setcounter{table}{0}
\renewcommand{\thetable}{ST\arabic{table}}
\setcounter{page}{1}

\begin{figure}[!hb]
\centering
\includegraphics[width=0.4\columnwidth]{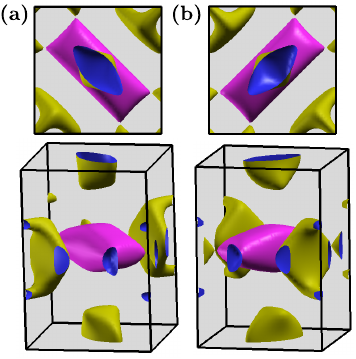}
\caption{
Top and overall view of the Fermi surface of RuO$_{2}$ (top and bottom panels, respectively) in the absence of spin--orbit coupling.
Results for spin-$\uparrow$ (a) and spin-$\downarrow$ (b) chanels.
\label{fig.el_fermi}
}
\end{figure}
\begin{figure}[!ht]
\centering
\includegraphics[width=\columnwidth]{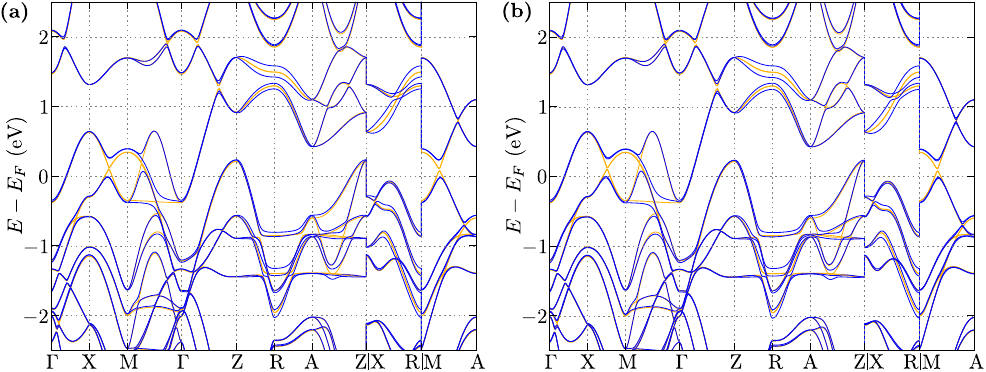}
\caption{
Electronic band structure of RuO$_{2}$ in the absence and in the presence of the spin--orbit coupling (orange and blue lines, respectively).
Results for ${\bm M} \parallel (101)$ (a) and ${\bm M} \parallel (111)$ (b).
\label{fig.el_band_tilted}
}
\end{figure}
\begin{figure}[!hb]
\centering
\includegraphics[width=\columnwidth]{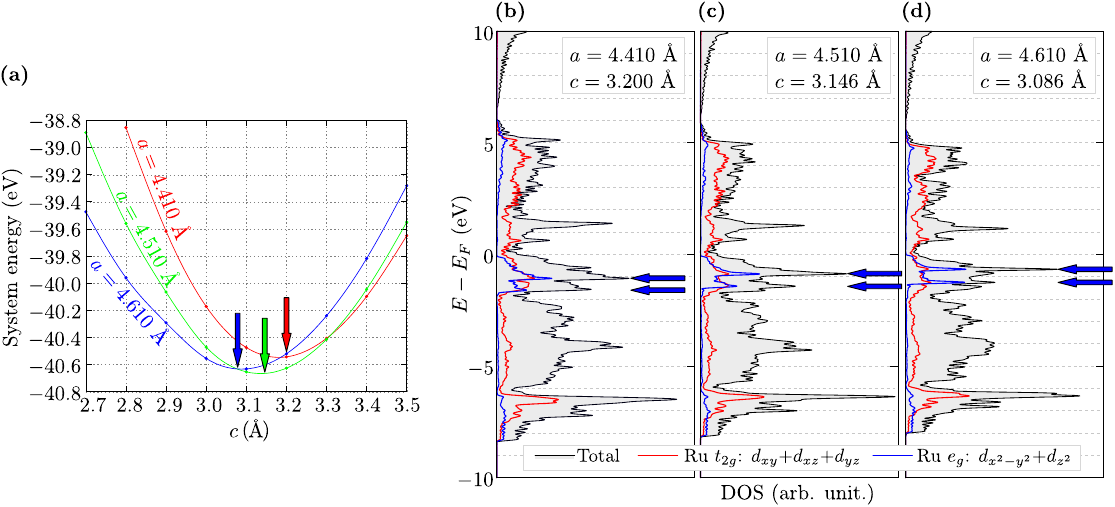}
\caption{
Influence of  the epitaxial strain on RuO$_{2}$ (in the absence of the spin--orbit coupling).
For fixed $a$ lattice constant (red, green, and blue lines), we find $c$ lattice parameter dependence of the system energy (a).
The optimized $c$ value can be found as the minimum of the system energy (marked by solid arrows).
Densities of states (DOS) for such optimized $c$ lattice parameters for different fixed $a$ lattice parameters (b)-(d).
Black line corresponds to the total DOS, while red and blue lines present partial DOSs related to Ru $t_{2g}$ and Ru $e_{g}$ states.
Blue solid arrows present positions of the peaks related to Ru $e_{g}$ states.
\label{fig.el_strain_dos}
}
\end{figure}

In this Supplemental Material, we present additional results:
\begin{itemize}
\item Table~\ref{tab.latt} -- The optimized lattice parameters within DFT+U scheme (lattice constants, Ru magnetic moments, and lattice free parameters describing O atoms positions).
\item Figure~\ref{fig.el_fermi} -- The Fermi surfaces for spin-$\uparrow$ and spin-$\downarrow$ electrons (in the absence of the spin--orbit coupling).
\item Figure~\ref{fig.el_band_tilted} -- The electronic band structures for Ru magnetic moments tilted from $c$ direction.
\item Figure~\ref{fig.el_strain_dos} -- Role of the epitaxial strain on the electronic density of states.
\end{itemize}

\begin{table*}[!hb]
\caption{
\label{tab.latt}
Lattice parameters of RuO$_{2}$ after optimization for different Hubbard-$U$ parameters. 
Experimentally reported values are: $a = 4.49$~\AA, and $c = 3.11$~\AA~\cite{zhu.strempfer.19}.
}
\begin{ruledtabular}
\begin{tabular}{cccccccc}
& $U$ (eV) & $a$ (\AA) & $c$ (\AA) & ${\bm M}_\text{Ru}$ ($\mu_{B}$) & $x_\text{O}$ & $y_\text{O}$ \\
\hline
& 0.0 & 4.522 & 3.119 & 0.000 & 0.8057 & 0.1943 & \\
& 0.5 & 4.518 & 3.119 & 0.025 & 0.8054 & 0.1946 & \\
& 1.0 & 4.516 & 3.120 & 0.062 & 0.8043 & 0.1957 & \\
& 1.5 & 4.524 & 3.123 & 0.928 & 0.8037 & 0.1963 & \\
& 2.0 & 4.533 & 3.124 & 1.169 & 0.8037 & 0.1963 & \\
& 2.5 & 4.541 & 3.123 & 1.338 & 0.8039 & 0.1961 & \\
& 3.0 & 4.544 & 3.123 & 1.442 & 0.8041 & 0.1959 & \\
& 3.5 & 4.545 & 3.123 & 1.516 & 0.8042 & 0.1958 & \\
& 4.0 & 4.544 & 3.122 & 1.580 & 0.8044 & 0.1956 & 
\end{tabular}
\end{ruledtabular}
\end{table*}

\end{document}